\documentclass[runningheads]{llncs}
\usepackage[T1]{fontenc}
\usepackage{graphicx}
\usepackage{tabularx}
\usepackage{subcaption}

\usepackage{ascmac}

\usepackage{amsmath}
\usepackage{amssymb}
\usepackage{amsfonts}

\newtheorem{assumption}{Assumption}

\newcommand\blfootnote[1]{%
  \begingroup
  \renewcommand\thefootnote{}\footnote{#1}%
  \addtocounter{footnote}{-1}%
  \endgroup
}
\begin{document}
\title{Modeling Hawkish-Dovish Latent Beliefs in Multi-Agent Debate-Based LLMs for Monetary Policy Decision Classification}

\titlerunning{Modeling Hawkish-Dovish Latent Beliefs}
%
\author{Kaito Takano\inst{1}\orcidID{0000-0000-0000-0000} \and
Masanori Hirano\inst{2}\orcidID{0000-0001-5883-8250} \and
Kei Nakagawa\inst{3}\orcidID{0000-0001-5046-8128}}
\authorrunning{K. Takano et al.}
%
\institute{Osaka Metropolitan University, Japan \email{takaito0423@gmail.com} \and Preferred Networks, Inc., Japan \email{research@mhirano.jp} \and Osaka Metropolitan University, Japan \email{kei.nak.0315@gmail.com}}

\maketitle              
\begin{abstract}
Accurately forecasting central bank policy decisions, particularly those of the Federal Open Market Committee~(FOMC) has become increasingly important amid heightened economic uncertainty. While prior studies have used monetary policy texts to predict rate changes, most rely on static classification models that overlook the deliberative nature of policymaking. This study proposes a novel framework that structurally imitates the FOMC's collective decision-making process by modeling multiple large language models~(LLMs) as interacting agents. Each agent begins with a distinct initial belief and produces a prediction based on both qualitative policy texts and quantitative macroeconomic indicators. Through iterative rounds, agents revise their predictions by observing the outputs of others, simulating deliberation and consensus formation. To enhance interpretability, we introduce a latent variable representing each agent’s underlying belief~(e.g., hawkish or dovish), and we theoretically demonstrate how this belief mediates the perception of input information and interaction dynamics. Empirical results show that this debate-based approach significantly outperforms standard LLMs-based baselines in prediction accuracy. Furthermore, the explicit modeling of beliefs provides insights into how individual perspectives and social influence shape collective policy forecasts.


\blfootnote{This paper does not reflect the view of the organizations the authors belong to. All errors in this paper are the responsibility of the authors.}
\keywords{Hawkish-Dovish  \and Multi-Agent LLMs \and Monetary Policy Decision \and FOMC}
\end{abstract}
\section{Introduction}

Monetary policy decision makings by central banks, especially changes in policy interest rates, have direct effects on financial markets and the cost of capital in the real economy~\cite{romer2000federal,nakamura2018high,jarocinski2018macroeconomic}.
Among central banks, the Federal Open Market Committee~(FOMC) of the U.S. is one of the most closely watched decision-making bodies in the world. Its policy rate decisions have significant influence not only on the U.S. economy but also on global financial markets. When the FOMC raises interest rates, U.S. dollar–denominated assets become more attractive, encouraging capital inflows into the United States. As a result, other countries—particularly emerging markets—often experience capital outflows, currency depreciation, and tighter financial conditions~\cite{kim2001international,bruno2015capital,couture2021financial}.

The FOMC consists of 12 voting members: 7 governors from the Federal Reserve Board and 5 presidents of regional Federal Reserve Banks. 
The committee meets eight times per year to determine the target range for the federal funds rate by majority vote. 
Each member has a unique background, shaped by their region and area of expertise, and thus brings a different beliefs to monetary policy. 
These beliefs are often categorized as either “dovish”~(favoring more accommodative policy) or “hawkish”~(favoring tighter policy).
The FOMC’s decision-making process is designed to reflect these diverse views. 
During each meeting, members present their policy beliefs, engage in debates, and ultimately reach a consensus. 
On the final day of the meeting, the committee releases a policy statement that announces the interest rate decision and outlines the economic reasoning behind it.\footnote{\url{https://www.federalreserve.gov/monetarypolicy/fomc.htm}}
To inform these decisions, the FOMC uses various types of information, including macroeconomic indicators such as the inflation rate. 
In addition, it consults the Beige Book, a qualitative report that reflects regional economic conditions. 
The Beige Book is published two weeks before each FOMC meeting and is compiled by the twelve regional Federal Reserve Banks. It summarizes findings from interviews with local businesses and stakeholders. Unlike typical economic data, the Beige Book is released in textual form and contains unstructured qualitative information.\footnote{\url{https://www.federalreserve.gov/monetarypolicy/publications/beige-book-default.htm}}
With the increasing complexity of economic conditions and the acceleration of inflation, accurately predicting both the timing and direction of monetary policy decisions has become a critical challenge for macroprudential policy design and investment strategy development  like~\cite{fulton2021forecasting,nakagawa2022inflation,brandao2024monetary}. 
In parallel, central banks themselves are increasingly adopting AI technologies. 
As monetary policy communication becomes more sophisticated and real-time analytics are further developed, external market participants are also expected to adopt more advanced analytical tools~\cite{fanta2024artificial,balsategui2024artificial}.
With this background, this study addresses the problem of predicting policy rate decisions made by the FOMC.

Many literature has attempted to forecast monetary policy decisions and market responses by analyzing the content of central bank communications, particularly unstructured policy texts such as the Beige Book and FOMC statements~\cite{hansen2018transparency,routledge2019machine,fujiwara2023treasury}. These studies typically focus on extracting features from the text such as tone, sentiment, or keyword frequencies and apply regression-based approaches to link them with policy outcomes or market variables.

However, existing approaches face two main limitations.
First, they fail to account for the deliberative decision-making process that is intrinsic to the institutional structure of the FOMC~\cite{schonhardt2013deliberating,lopez2016sequential}. 
As described earlier, FOMC members express divergent policy beliefs informed by differing economic perspectives, and these views are gradually reconciled through discussion and debate to form a final consensus. 
Yet, many prior studies treat the FOMC as a single unified decision-making entity, neglecting the internal dynamics and diversity of beliefs within the committee.

Second, most existing methods rely on static text analysis based on dictionaries or single-model predictions~\cite{hansen2018transparency,routledge2019machine,fujiwara2023treasury}.
As a result, they struggle to integrate macroeconomic indicators with textual information in a cohesive manner, and they lack the capacity to model how individual beliefs emerge and evolve among committee members during the policy formation process.

To address the aforementioned challenges, this study proposes a novel descriptive framework for policy rate determination that structurally simulates the deliberative decision-making process of the FOMC. 
The proposed framework models the process using multiple pre-trained LLMs, each acting as an autonomous agent. 
Each LLM-based agent represents an FOMC member with a distinct policy beliefs—such as dovish or hawkish—and operates independently.
Each agent receives as input a combination of qualitative information and quantitative macroeconomic indicators. 
Based on this input, the agent generates an initial policy decision. 
Then, in successive rounds, each agent sequentially observes the decisions made by other agents and updates its own output accordingly. 
In each round, agents revise their decisions by incorporating the judgments of others as part of an ongoing debate. 
This iterative, interdependent process is designed to structurally mimic the institutional dynamics of actual FOMC meetings, namely, the expression of beliefs, debate, beliefs adjustment, and eventual consensus formation.

Furthermore, following~\cite{estornell2024multi}, we explicitly model each agent’s internal policy beliefs as a discrete latent variable. We assume that the policy label is generated probabilistically based on this latent stance, and we formalize the decision process as a Bayesian generative model. 
Theoretically, the output of each agent is generated as a function of (i) the input data (text and macroeconomic indicators), (ii) the observed decisions of other agents, and (iii) the agent's own latent belief, serving as a mediating variable. This structure can capture the interaction between external evidence, peer influence, and individual predispositions.

The goal of this study is to answer two research questions using the proposed structural framework.
Although debate-based multi-agent LLMs have proven effective in enhancing reasoning, accuracy, and consensus-building~\cite{du2023improving,chan2024chateval,liang2024encouraging}, there is no existing research that applies structured debate-based LLM frameworks specifically to the financial domain.

First, we empirically evaluate whether the proposed model can replicate actual FOMC policy decisions. Specifically, we test whether the model, given the Beige Book and macroeconomic indicators available before each FOMC meeting, can produce the same decision outcome as the actual FOMC. This serves as an empirical validation of whether our model, which incorporates the institutional features of the FOMC, also possesses practical predictive power.

Second, we aim to identify which components of the proposed framework are most critical for decision-making. To this end, we conduct an ablation study to quantitatively assess the contribution of various components—including textual information, macroeconomic variables, peer agent outputs, and latent beliefs variables. This analysis reveals the key informational and structural factors that drive interest rate decisions in our framework.

\section{Related Work}
Recent advances in LLMs have been remarkable. Notably, cutting-edge models such as ChatGPT \cite{chatgpt}, GPT-4 \cite{GPT4}, Claude, and Gemini have demonstrated performance and generalizability far beyond previous models, finding applications across diverse fields. These models trace their origins to the Transformer architecture \cite{Vaswani2017}, evolving through subsequent developments including BERT \cite{Devlin2018} and the GPT series \cite{GPT-1,GPT-2,GPT-3}, ultimately leading to the emergence of LLMs.

Since LLMs are trained on diverse texts, they acquire thought patterns that people have.
This capability has been leveraged in research applications such as opinion research, which can generate responses specific to demographic characteristics or individual preferences \cite{qu2024performance,gatto2024scope}.

Furthermore, Park {\it et al.} \cite{park2023generative} constructed an RPG-style social simulation featuring LLM-based agents with various roles to investigate how new social structures form. Additionally, ResearchTown \cite{yu2024researchtown} - an LLM-based research community simulation environment - was developed to model and analyze the emergence of new academic papers through collaborative research among scholars. Takata {\it et al.} \cite{takata2024spontaneous} developed an LLM-based multi-agent environment to examine the mechanisms underlying the manifestation of individual characteristics.

Building upon these studies, this study investigates whether LLMs can reproduce the process of collective decision-making in the financial domain.

It is well established that when multiple decision-making entities interact, complex behaviors emerge rather than linear patterns.
Notable studies include those by Schelling \cite{schelling1969,schelling1969} and Axelrod \cite{axelrod1997dissemination}, which demonstrate that when numerous agents with extremely simple decision-making behaviors interact, phenomena such as segregation and cultural diversity can emerge.
Similarly, in financial markets, Lux {\it et al.} \cite{Lux1999} demonstrated that agent interactions are crucial for reproducing stylized facts in financial market simulations.

Given those backgrounds, this study aims to reproduce the decision-making process by simulating interactions among members of the FOMC using LLMs, with the goal of reproducing the complex collective decision-making behavior of the members.

\section{Problem Formulation}
In this study, we consider a three-class classification problem to predict the central bank’s next policy rate decision~(Raise, Hold, or Lower) based on financial texts and associated numerical macroeconomic data.

Unlike conventional supervised learning approaches that optimize model parameters using labeled training data, our approach relies on LLMs that have already been pre-trained. These LLMs are used directly for prediction without additional parameter tuning.

Let $x$ denote a financial text, and let $v \in \mathbb{R}^{d}$ denote a vector of relevant numerical indicators. Here, $x$ is a natural language document, whose length and structure can vary. The vector $v$ summarizes $d$-dimensional quantitative macroeconomic indicators relevant to policy decisions, such as recent inflation or unemployment rates.

The corresponding label $y$ is an element of the discrete set $\mathcal{Y}$ representing the policy rate decision:
\begin{align}    
    \mathcal{Y} = \{\text{Raise},\,\text{Hold},\,\text{Lower}\}
\end{align}

Our dataset consists of tuples of past published texts, the numerical indicators available at the time, and the actual policy decision made by the central bank. However, we do not fine-tune the model using these data.

Instead, each LLM receives the input pair $(x, v)$ as a prompt and generates a prediction $z \in \mathcal{Y}$ based on its fixed, pre-trained parameters $\phi$:
\begin{align}\label{eq:pres_extended}
    z \sim P_{\mathrm{LLM}}\bigl(z \mid x,\,v,\,\phi\bigr)
\end{align}

This formulation treats the LLM as a probabilistic predictor over the set of possible policy actions, conditioned on both textual and numerical inputs.

\section{Proposed Method}

In this study, we propose a framework in which a total of $n$ LLMs act as agents and conduct a multi-round debate over $T$ iterations to reach a collective decision for classification.

Each agent $i$ has pre-trained parameters $\phi_i$ and receives as input a financial text $x$ and numerical vector $v$. At the initial round $t=0$, agent $i$ independently generates a policy label based on its own model as follows:
\begin{align}
    z_{i}^{(0)} \sim P_{\mathrm{LLM}}\bigl(z \mid x,\,v,\,\phi_{i}\bigr),
\qquad z_{i}^{(0)} \in \mathcal{Y}.
\end{align}
Here, $z_{i}^{(0)}$ represents the initial label proposed by agent $i$. The distribution $P_{\mathrm{LLM}}(\cdot \mid x, v, \phi_i)$ is fully determined by the fixed pre-trained parameters $\phi_i$ of agent $i$.

From round $t \ge 1$ onward, each agent observes the set of predictions made by all agents in the previous round:
\begin{align}
    Z^{(t-1)} = \bigl\{\,z_{1}^{(t-1)},\,z_{2}^{(t-1)},\,\dots,\,z_{n}^{(t-1)}\bigr\},
\end{align}
and updates its prediction accordingly:
\begin{align} \label{main}
    z_{i}^{(t)} \sim P_{\mathrm{LLM}}\bigl(z \mid x,\,v,\,Z^{(t-1)},\,\phi_{i}\bigr),
\qquad z_{i}^{(t)} \in \mathcal{Y}.
\end{align}
The conditional distribution $P_{\mathrm{LLM}}(\cdot \mid x,\,v,\,Z^{(t-1)},\,\phi_i)$ represents the agent's output distribution given not only the original inputs $(x, v)$, but also the collective output of all agents in the previous round $Z^{(t-1)}$ as additional context.

For instance, agent $i$ may incorporate the opinions of others along with updated macroeconomic indicators such as inflation and unemployment into its prompt. This allows the agent to refine its judgment by comparing its beliefs with others and the latest data.

This iterative process forms a debate-based decision-making framework, where each agent updates its output in each round by considering both the opinions of others and the economic data.

At the final round $T$, let $a(\cdot)$ denote a label extraction function that maps each agent's response to a canonical decision. If all agents agree on the same label, i.e.,
\begin{align}
    a\bigl(z_{1}^{(T)}\bigr) = a\bigl(z_{2}^{(T)}\bigr) = \cdots = a\bigl(z_{n}^{(T)}\bigr),
\end{align}
then the shared label $z = a(z_{i}^{(T)})$ is adopted as the final prediction, and the debate terminates. If consensus is not achieved after the maximum number of rounds $T$, the label with the highest number of supporting agents is selected as the final output.

\subsection{Latent Policy Beliefs}

While the debate-based framework described above enables iterative opinion exchange among agents, it lacks transparency in explaining why each agent selects a particular label, that is, the internal policy beliefs remains a black box. 
To address this, we introduce an explicit latent variable representing the agent's underlying policy beliefs, following the framework proposed by \cite{estornell2024multi}. 
This latent variable, referred to as a hawk-dove stance, governs how each agent interprets the input text $x$ and numerical data $v$.

We define a discrete latent space of policy stances as:
\begin{align}
    \Theta = \{\theta_{1},\,\theta_{2},\,\dots,\,\theta_{K}\},
\end{align}
where each element $\theta_k$ represents a cluster of policy beliefs, such as strongly hawkish (favoring aggressive rate hikes), moderately hawkish, or dovish (favoring monetary easing). 
The cardinality $K = |\Theta|$ denotes the number of distinct stance categories.

Suppose agent $i$ is given the input text $x$, numerical data $v$, and the set of labels from the previous round,
\begin{align}
    Z^{(t)} = \{z_{1}^{(t)},\,z_{2}^{(t)},\,\dots,\,z_{n}^{(t)}\},
\end{align}
and outputs a label in round $t+1$. The probability of its response is modeled as:
\begin{align}\label{eq:outputprob_extended}
    P_{\mathrm{LLM}}\bigl(z_{i}^{(t+1)} \mid x,\,v,\,Z^{(t)},\,\phi_{i}\bigr),
\end{align}
which can be expanded via marginalization over the latent beliefs variable $\theta \in \Theta$ as follows.

\begin{lemma}[Latent Beliefs Decomposition]
\label{lemma1_extended}
\begin{align}
    &P_{\mathrm{LLM}}\bigl(z_{i}^{(t+1)} \mid x,\,v,\,Z^{(t)},\,\phi_{i}\bigr)\nonumber\\
    &~~=\sum_{\theta \in \Theta}
P\bigl(z_{i}^{(t+1)} \mid \theta,\,x,\,v,\,Z^{(t)},\,\phi_{i}\bigr)\,P\bigl(\theta \mid x,\,v,\,Z^{(t)},\,\phi_{i}\bigr).
\end{align}
Here, $P(z_{i}^{(t+1)} \mid \theta,\,x,\,v,\,Z^{(t)},\,\phi_{i})$ represents the conditional probability that agent $i$ outputs label $z_{i}^{(t+1)}$ given a known stance $\theta$, observed inputs $x$, $v$, and previous agent responses $Z^{(t)}$. The posterior distribution $P(\theta \mid x,\,v,\,Z^{(t)},\,\phi_{i})$ captures the probability that agent $i$ adopts belief $\theta$ after observing $x$, $v$, and $Z^{(t)}$.
\end{lemma}

\begin{assumption}[Conditional Independence Given Latent Belief]
For any round $t$ and any belief $\theta \in \Theta$, the output $z_{i}^{(t+1)}$ by agent $i$ is conditionally independent of $x$, $v$, and $Z^{(t)}$ once $\theta$ is known:
\begin{align}
\label{eq:assumption_extended}
P\bigl(z_{i}^{(t+1)} \mid \theta,\,x,\,v,\,Z^{(t)},\,\phi_{i}\bigr)
=
P\bigl(z_{i}^{(t+1)} \mid \theta,\,\phi_{i}\bigr)
\quad
(\forall\,\theta\in\Theta,\;t\ge0).
\end{align}
\end{assumption}

This assumption implies that once the belief $\theta$ is fixed, the final label output becomes independent of the input text $x$, numerical data $v$, and others' responses.

Under this assumption, the output probability can be simplified as follows.

\begin{lemma}[Posterior Decomposition]
\label{lemma2_extended}
Given Assumption~\ref{eq:assumption_extended}, the output probability in Equation~\eqref{eq:outputprob_extended} reduces to:
\begin{align}
    P_{\mathrm{LLM}}\bigl(z_{i}^{(t+1)} \mid x,\,v,\,Z^{(t)},\,\phi_{i}\bigr)
    = \sum_{\theta \in \Theta}
    P\bigl(z_{i}^{(t+1)} \mid \theta,\,\phi_{i}\bigr)
    \,P\bigl(\theta \mid x,\,v,\,Z^{(t)},\,\phi_{i}\bigr).
\end{align}
Moreover, the posterior distribution over stances is given by:
\begin{align}    
\label{eq:lemma42_posterior_extended}
P\bigl(\theta \mid x,\,v,\,Z^{(t)},\,\phi_{i}\bigr)
\;\propto\;
P\bigl(x,\,v \mid \theta,\,\phi_{i}\bigr)\,
P\bigl(\theta \mid \phi_{i}\bigr)\,
\prod_{j=1}^{n}
P\bigl(z_{j}^{(t)} \mid \theta,\,\phi_{i}\bigr).
\end{align}
Here, $P(z_{j}^{(t)} \mid \theta,\,\phi_{i})$ is the probability that agent $j$ produces label $z_{j}^{(t)}$ given stance $\theta$. The term $P(x, v \mid \theta, \phi_i)$ is the likelihood of observing $(x, v)$ under belief $\theta$, and $P(\theta \mid \phi_i)$ represents the prior belief distribution of agent $i$. 
The product term reflects how well the previous responses of other agents support belief $\theta$.
\end{lemma}

Using this framework, each agent determines its own belief $\theta$ by jointly considering the input text $x$, numerical indicators $v$, and the previous outputs from other agents. Once a belief $\theta$ is selected, the agent generates its final label conditioned on that belief. 
These lemmas provide a formal foundation for modeling the internal reasoning process of each agent. Specifically, they describe how an agent interprets input signals, selects a monetary policy belief, and generates a final decision in a mathematically consistent manner.

\section{Experiment}
In this section, we conduct two experiments to evaluate the effectiveness of our framework.  
First, we test whether our model can reproduce actual FOMC policy decisions (Raise, Hold, or Lower) using the Beige Book and macroeconomic indicators available before each meeting.  
Second, we perform an ablation study to analyze the contribution of each component in our model such as textual information, numerical indicators, peer predictions, and belief variables to overall prediction performance.

\subsection{Dataset}
Our empirical analysis spans all scheduled FOMC meetings from January 2000 to December 2025.  
The next subsections describe three sources of information that match the inputs used by our agents.

\subsubsection{Policy Rate:}
As our target variable, we use the actual federal funds policy rate decision made by the FOMC at each meeting.
We classify each decision into one of three categories: Raise, Hold, or Lower, based on changes in the target range of the policy rate compared to the previous meeting.
We also use as an input to the LLM, since information on the current level and trend of the policy rate may be important in decision the policy rate.
We obtain this data from the Bloomberg terminal\footnote{\url{https://www.bloomberg.com/professional/products/bloomberg-terminal/}}.

\subsubsection{Macroeconomic Indicators:}
To construct the input data for our model, we use both macroeconomic indicators and qualitative policy texts.  
In accordance with the Federal Reserve's dual mandate: maximum employment and price stability, we use the unemployment rate and the inflation rate as macroeconomic indicators.  
We obtain these data from the Federal Reserve Economic Data~(FRED) database\footnote{\url{https://fred.stlouisfed.org/}}.

\subsubsection{Beige Book:}
The Beige Book compiles economic conditions of the jurisdictions of each regional Federal Reserve Bank, and is published two weeks prior to the FOMC meetings. 
Its topics span a wide range, including GDP, inflation, employment, manufacturing, agriculture, tourism, real estate, and more. Given its role as material for debate at FOMC meetings, the Beige Book also serves as one of the resources for speculating on the outcomes of the next FOMC meeting\cite{fujiwara2023treasury}. 

In our research, we work with the Beige Book corpus created by \cite{KaitoTakano2023frb}. 
This dataset is structured to be user-friendly for a range of analytical purposes, reflecting insights from existing studies. 
Its design promotes widespread use, allowing for consistent comparisons in various experiments.
Each sentence in this dataset is assigned a topic, and we use 'overall economic activity' or 'summary' topic sentences.

\subsection{Experimental Settings}
Given the FOMC's eight meetings per year, our study's experimental period encompasses approximately 200 distinct time slices.
We randomly select these slices such that 15 instances correspond to "Raise", 30 to "Hold", and 15 to "Lower" decisions.
However, periods with minimal policy rate changes (e.g., between 2009-2015) or consecutive rate hikes are considered too trivial as prediction tasks, so we exclude any slices where the true policy decision matches either the preceding or following decision.

We consider the task of predicting the FOMC's policy rate decision (Raise, Hold, or Lower) at the time of the Beige Book release, with policy decisions to be made two weeks later.
The input prompt includes: Beige Book textual data, economic indicators from the past three months, and the last two policy rate decisions.

We set the number of agents $n$ to 7 to facilitate majority voting.
Each agent is assigned a specific belief profile, consisting of: one each of "Strong Hawkish", "Moderately Hawkish", "Moderately Dovish", "Strong Dovish", and three "Neutral" agents.
Detailed definitions of these beliefs are presented in Table \ref{tab:belief}.

\begin{table}
\centering
\caption{Belief Description}
\label{tab:belief}
\scalebox{0.70}{
\begin{tabular}{l|l}
\hline
Belief & Description \\
\hline
Strong Hawkish & Prioritizes controlling inflation and supports aggressive interest rate hikes \\
Moderately Hawkish & Proposes tightening of inflation but is mindful of economic downturns \\
Neutral & Makes careful decisions while monitoring the balance between prices and the economy \\
Moderately Dovish & Emphasizes supporting the economy while also paying a certain amount of attention to prices \\
Strong Dovish & Prioritizes economic recovery and actively supports interest rate cuts \\
\hline
\end{tabular}
}
\end{table}
For our LLM implementation, we will use GPT-4o-mini\footnote{gpt-4o-mini-2024-07-18}, with API access provided by OpenAI\footnote{\url{https://openai.com/}}.
The temperature hyperparameter, which controls the randomness in generation, is set to 1 to enable diverse debate patterns, while all other hyperparameters are configured to their default values.
To ensure consistent output format, we will employ Structured Outputs\footnote{\url{https://platform.openai.com/docs/guides/structured-outputs?api-mode=responses}}. This configuration specifies that the output should include both a 'label' and its supporting 'justification'.

The maximum number of rounds is set to 10 to allow sufficient debate time. However, debate will terminate immediately when all LLMs reach complete agreement on their outputs.

The actual prompt used is as follows:
\begin{itembox}[l]{\textbf{Round $t=0$}}
\small
Today is \{\underline{Month}\}. \\
You will be given beige book text data, associated macroeconomic numerical data, historical policy rate, and a prior belief of central bank policy. \\
Based on these inputs, predict whether the central bank will Raise, Hold, or Lower the policy rate after two weeks. You should provide a brief justification for your answer, and you must output one of the three labels: Raise, Hold, or Lower. \\
Please also note that policy rate changes should be implemented with appropriate speed, and that taking Hold is not necessarily always the best approach. \\
Belief: \{\underline{$\mathrm{Belief}_k$}\} \\
Beige Book Text Data: \{\underline{Text}\} \\
Macroeconomic Numerical Data: \{\underline{Indicators}\} \\
Historical Policy Rate: \{\underline{Rates}\}
\end{itembox}

\begin{itembox}[l]{\textbf{Round $t>0$}}
\small
Today is \{\underline{Month}\}. \\
Several other models have already given their predictions and current beliefs: \\
$\mathrm{Model}_1$: Label is \{\underline{$\mathrm{Prediction}_{1}^{(t-1)}$}\}. \{\underline{$\mathrm{Justification}_{1}^{(t-1)}$}\} (\{\underline{$\mathrm{Belief}_1$}\}) \\
$\mathrm{Model}_2$: Label is \{\underline{$\mathrm{Prediction}_{2}^{(t-1)}$}\}. \{\underline{$\mathrm{Justification}_{2}^{(t-1)}$}\} (\{\underline{$\mathrm{Belief}_2$}\}) \\
$\cdot\cdot\cdot$ \\
$\mathrm{Model}_n$: Label is \{\underline{$\mathrm{Prediction}_{n}^{(t-1)}$}\}. \{\underline{$\mathrm{Justification}_{n}^{(t-1)}$}\} (\{\underline{$\mathrm{Belief}_n$}\}) \\
\\
Now you should consider these responses and beliefs. \\
You are again given beige book text data, associated macroeconomic numerical data, historical policy rate, your current prediction, and your current belief. \\
Use all of these to predict whether the central bank will Raise, Hold, or Lower the policy rate after two weeks. You should provide a brief justification for your answer, and you must output one of the three labels: Raise, Hold, or Lower. \\
Please also note that policy rate changes should be implemented with appropriate speed, and that taking Hold is not necessarily always the best approach. \\
Belief: \{\underline{$\mathrm{Belief}_k$}\} \\
Current Prediction:  \{\underline{$\mathrm{Prediction}_{k}^{(t-1)}$}\} \\
Beige Book Text Data: \{\underline{Text}\} \\
Macroeconomic Numerical Data: \{\underline{Indicators}\} \\
Historical Policy Rate: \{\underline{Rates}\}
\end{itembox}

Here, {underline} represents a variable containing information embedded in the prompt.
The correspondence with Equation \ref{main} is as follows:
\begin{description}
   \item[$z_{i}^{(t)} = (\mathrm{Prediction}_{i}^{(t)}, \mathrm{Justification}_{i}^{(t)})$ : ] Predicted Label and Its Justification
   \item[$\phi_{i} = \mathrm{Belief}_{i}$ : ] Belief
   \item[$x = \mathrm{Text}$ : ] Beige Book Text Data
   \item[$v = \mathrm{Indicators},\mathrm{Rates}$ : ] Macroeconomic Numerical Data
\end{description}

The above experiments will be referred to as experiment (1) in the following.

\subsection{Ablation Study}
As part of our ablation study, we will investigate the impact of: (2) textual information, (3) macroeconomic indicators, and (4) the current policy rate level on prediction accuracy.
The prompt structure remains largely unchanged from Experiment (1), with only the following single sentence varying: 'You will be given beige book text data, associated macroeconomic numerical data, historical policy rate, and a prior belief of central bank policy.' We will modify this information content to assess its influence.
For example, the prompt without textual information would appear as follows:
\begin{itembox}[l]{\textbf{Round $t=0$ (Remove Text)}}
\small
Today is \{\underline{Month}\}. \\
You will be given beige associated macroeconomic numerical data, historical policy rate, and a prior belief of central bank policy. \\
Based on these inputs, predict whether the central bank will Raise, Hold, or Lower the policy rate after two weeks. You should provide a brief justification for your answer, and you must output one of the three labels: Raise, Hold, or Lower. \\
Please also note that policy rate changes should be implemented with appropriate speed, and that taking Hold is not necessarily always the best approach. \\
Belief: \{\underline{$\mathrm{Belief}_k$}\} \\
Macroeconomic Numerical Data: \{\underline{Indicators}\} \\
Historical Policy Rate: \{\underline{Rates}\}
\end{itembox}

\begin{itembox}[l]{\textbf{Round $t>0$ (Remove Text)}}
\small
Today is \{\underline{Month}\}. \\
Several other models have already given their predictions and current beliefs: \\
$\mathrm{Model}_1$: Label is \{\underline{$\mathrm{Prediction}_{1}^{(t-1)}$}\}. \{\underline{$\mathrm{Justification}_{1}^{(t-1)}$}\} (\{\underline{$\mathrm{Belief}_1$}\}) \\
$\mathrm{Model}_2$: Label is \{\underline{$\mathrm{Prediction}_{2}^{(t-1)}$}\}. \{\underline{$\mathrm{Justification}_{2}^{(t-1)}$}\} (\{\underline{$\mathrm{Belief}_2$}\}) \\
$\cdot\cdot\cdot$ \\
$\mathrm{Model}_n$: Label is \{\underline{$\mathrm{Prediction}_{n}^{(t-1)}$}\}. \{\underline{$\mathrm{Justification}_{n}^{(t-1)}$}\} (\{\underline{$\mathrm{Belief}_n$}\}) \\
\\
Now you should consider these responses and beliefs. \\
You are again given associated macroeconomic numerical data, historical policy rate, your current prediction, and your current belief. \\
Use all of these to predict whether the central bank will Raise, Hold, or Lower the policy rate after two weeks. You should provide a brief justification for your answer, and you must output one of the three labels: Raise, Hold, or Lower. \\
Please also note that policy rate changes should be implemented with appropriate speed, and that taking Hold is not necessarily always the best approach. \\
Belief: \{\underline{$\mathrm{Belief}_k$}\} \\
Current Prediction:  \{\underline{$\mathrm{Prediction}_{k}^{(t-1)}$}\} \\
Macroeconomic Numerical Data: \{\underline{Indicators}\} \\
Historical Policy Rate: \{\underline{Rates}\}
\end{itembox}

Furthermore, to examine the predictive capability of a simple approach without beliefs (5), we will investigate accuracy using the following prompt.
To maintain consistent conditions, we will perform seven predictions and determine policy decisions through majority voting.
\begin{itembox}[l]{\textbf{Round $t=0$ (Remove Belief)}}
\small
Today is \{\underline{Month}\}. \\
You will be given beige book text data, associated macroeconomic numerical data and historical policy rate. \\
Based on these inputs, predict whether the central bank will Raise, Hold, or Lower the policy rate after two weeks. You should provide a brief justification for your answer, and you must output one of the three labels: Raise, Hold, or Lower. \\
Please also note that policy rate changes should be implemented with appropriate speed, and that taking Hold is not necessarily always the best approach. \\
Beige Book Text Data: \{\underline{Text}\} \\
Macroeconomic Numerical Data: \{\underline{Indicators}\} \\
Historical Policy Rate: \{\underline{Rates}\}
\end{itembox}

Finally, to demonstrate the utility of multi-round agent debate in reaching consensus, we define Experiment (6) as the majority vote result from round 0 of the main experiment (1).

\subsection{Results \& Discussion}
We conducted experiments (1) through (6), calculating three evaluation metrics—Precision, Recall, and F1-Score using macro averaging. The results are presented in Table \ref{tab:results}.

\begin{table*}[h]
    \centering
    \caption{All results (Precision, Recall, F1-Score)}
    \label{tab:results}
    \scalebox{1}{
    \begin{tabular}{l|r|r|r} 
        \hline
        Experimental Settings & Precision & Recall & F1-Score \\ 
        \hline 
        \hline
        (1) Proposed Method & \textbf{0.549} & \textbf{0.467} & \textbf{0.476} \\ 
        (2) Remove Beige Book & 0.399 & 0.422 & 0.385 \\ 
        (4) Remove Historical Policy Rate & 0.535 & \underline{0.456} & \underline{0.464} \\ 
        (5) Remove Belief & 0.514 & 0.411 & 0.399 \\ 
        (6) No Debate & \underline{0.543} & 0.422 & 0.415 \\ 
        \hline
    \end{tabular}
    }
\end{table*}

Experimental results showed that (1) yielded the best performance.
The confusion matrix for (1) is presented in Table \ref{tab:confusion_matrix}.
\begin{table*}[h]
    \centering
    \caption{Confusion Matrix}
    \label{tab:confusion_matrix}
    \scalebox{1}{
    \begin{tabular}{rl|ccc} 
        \hline
        & &  & Predicted &  \\ 
        & & Raise & Hold & Lower \\ 
        \hline 
        \hline
         & Raise & 7 & 8 & 0 \\ 
        Actual & Hold & 8 & 20 & 2 \\ 
         & Lower & 0 & 11 & 4 \\ 
        \hline
    \end{tabular}
    }
\end{table*}

As shown in Table \ref{tab:confusion_matrix}, no significant directional errors were observed—for instance, raising rates when lowering was the actual decision or vice versa.
The information sources used in this research demonstrate that policy direction can be reasonably constrained.

Results from Experiment (2) indicate that Beige Book information proves valuable for policy rate determination.
The Beige Book contains descriptions of overall U.S. price and employment conditions, which align with the Fed's Dual Mandate.
Furthermore, results from Experiment (3) confirm that both quantitative and textual macroeconomic indicators are crucial for policy rate decision-making.
However, the presence of recent policy rate trends (Experiment (4)) showed no significant impact on prediction accuracy.
This finding may be partially explained by our intentional selection of policy rate transition points.
When policy rate trends are present, it could potentially be useful for predicting rate changes, but excessive reliance on current trend biases may pose risks and requires caution.

Experiments (5) and (6) represent majority voting results without inter-agent debate.
These results demonstrate that the iterative decision-making approach—where each agent refines its judgment based on others' opinions over multiple rounds—is an effective methodology.

Based on the empirical results from Experiments (1) and (6), we present in Table \ref{tab:count}: (1) aggregated policy decisions for each belief category after the final round, and (6) aggregated policy decisions for each belief category before any debate.

\begin{table*}[h]
    \centering
    \caption{Aggregate policy decisions by belief category}
    \label{tab:count}
    \scalebox{1}{
    \begin{tabular}{l|ccc||ccc} 
        \hline
        &  & (1)After debate &  &  & (6)Before debate &  \\ 
        & Raise & Hold & Lower & Raise & Hold & Lower \\ 
        \hline 
        \hline
        Strong Hawkish & 30 & 29 & 1 & 33 & 22 & 5 \\ 
        Moderately Hawkish & 28 & 32 & 0 & 27 & 27 & 6 \\ 
        Neutral & 31 & 136 & 13 & 45 & 117 & 18 \\ 
        Moderately Dovish & 3 & 53 & 4 & 15 & 36 & 9 \\ 
        Strong Dovish & 4 & 46 & 10 & 15 & 32 & 13 \\ 
        \hline
        Total & 96 & 296 & 28 & 135 & 234 & 51 \\ 
        \hline
    \end{tabular}
    }
\end{table*}

Furthermore, the aggregated policy decision counts for "Raise", "Hold", and "Lower" in Experiment (5) without belief information are 111, 280, and 29 respectively.
From Experiment (5) results, we observe that regardless of belief type, there is a consistent predictive bias toward "Hold", followed by "Raise" and then "Lower".
This bias persists even after incorporating belief information.
Comparing Experiment (5) with (6), we note that providing belief information somewhat mitigates the tendency toward "Hold" dominance, resulting in more diverse policy decision outcomes.
By allowing each agent to generate diverse opinions based on initial beliefs in Experiment (1) and then facilitating their mutual exchange, we achieve the optimal results shown in \ref{tab:results}.
Figure \ref{tab:transition_matrix} illustrates the policy decision changes resulting from the debate process.

\begin{table*}[h]
    \centering
    \caption{Transition Matrix}
    \label{tab:transition_matrix}
    \scalebox{1}{
    \begin{tabular}{rl|ccc} 
        \hline
        & &  & (6)Before debate &  \\ 
        & & Raise & Hold & Lower \\ 
        \hline 
        \hline
         & Raise & 80 & 55 & 0 \\ 
        (1)After debate & Hold & 16 & 217 & 1 \\ 
         & Lower & 0 & 24 & 27 \\ 
        \hline
    \end{tabular}
    }
\end{table*}

\section{Conclusion}
In this study, we proposed a structured modeling approach that simulates the FOMC's collective decision-making process using multiple large language models.  
Each agent integrates policy texts and macroeconomic indicators, updates its belief through debate, and makes a final prediction.  
We also introduced a latent belief variable and theoretically showed that it mediates the relationship between input information and the agent’s decision, thereby enhancing the interpretability of agent behavior.
We evaluated the method on 60 meetings held between 2000 and 2025. The full model that includes both debate and beliefs reached an F1 score of 0.48, outperforming versions that remove the debate, the beliefs, the Beige Book, or the macroeconomic indicators. The ablation study showed that the Beige Book is especially important for accuracy, and that the debate rounds lessen the strong Hold bias seen when the agents do not interact.
Several limitations remain in this study: The belief space is small and discrete, the framework relies on a single language model family, and the experiments cover only the FOMC. For further study, we will test continuous belief spaces, introduce safeguards against hallucination, and apply the framework to other central bank policy committees such as the European Central Bank and the Bank of Japan.

\appendix
\section{Proof of Lemma}
\subsection{Proof of Lemma~\ref{lemma1_extended}}
\begin{proof}
Using the law of total probability, we can marginalize over the latent policy belief $\theta$ to express the label generation probability at round $t+1$ as:
\begin{align*}
P_{\mathrm{LLM}}\bigl(z_{i}^{(t+1)} \mid x,\,v,\,Z^{(t)},\,\phi_{i}\bigr)
&=\sum_{\theta \in \Theta}
P\bigl(z_{i}^{(t+1)},\,\theta \mid x,\,v,\,Z^{(t)},\,\phi_{i}\bigr) \\
&=\sum_{\theta \in \Theta}
P\bigl(z_{i}^{(t+1)} \mid \theta,\,x,\,v,\,Z^{(t)},\,\phi_{i}\bigr)\,
P\bigl(\theta \mid x,\,v,\,Z^{(t)},\,\phi_{i}\bigr).
\end{align*}
Here, $P(z_{i}^{(t+1)} \mid \theta,\,x,\,v,\,Z^{(t)},\,\phi_{i})$ denotes the conditional probability that agent $i$ generates label $z_{i}^{(t+1)}$ given a known latent belief $\theta$ and observations $x$, $v$, and $Z^{(t)}$. 
The term $P(\theta \mid x,\,v,\,Z^{(t)},\,\phi_{i})$ represents the posterior distribution over stances after observing $x$, $v$, and $Z^{(t)}$. This completes the proof of Lemma~\ref{lemma1_extended}.
\end{proof}

\subsection{Proof of Lemma~\ref{lemma2_extended}}
\begin{proof}
According to Assumption~\ref{eq:assumption_extended}, we have:
\begin{align}
    P\bigl(z_{i}^{(t+1)} \mid \theta,\,x,\,v,\,Z^{(t)},\,\phi_{i}\bigr)=P\bigl(z_{i}^{(t+1)} \mid \theta,\,\phi_{i}\bigr)\quad(\forall\,\theta\in\Theta,\;t\ge0).
\end{align}
Substituting this into the decomposition from Lemma~\ref{lemma1_extended}, we obtain:
\begin{align}
P_{\mathrm{LLM}}\bigl(z_{i}^{(t+1)} \mid x,\,v,\,Z^{(t)},\,\phi_{i}\bigr)
&=
\sum_{\theta \in \Theta}
P\bigl(z_{i}^{(t+1)} \mid \theta,\,\phi_{i}\bigr)\,
P\bigl(\theta \mid x,\,v,\,Z^{(t)},\,\phi_{i}\bigr),
\label{eq:proof_extended_step1}
\end{align}
which corresponds to the first part of Lemma~\ref{lemma2_extended}.

Next, we derive the posterior distribution $P\bigl(\theta \mid x,\,v,\,Z^{(t)},\,\phi_{i}\bigr)$ using Bayes' theorem:
\begin{equation}
\label{eq:proof_extended_posterior_start}
P\bigl(\theta \mid x,\,v,\,Z^{(t)},\,\phi_{i}\bigr)
=
\frac{P\bigl(x,\,v,\,Z^{(t)} \mid \theta,\,\phi_{i}\bigr)\,
      P\bigl(\theta \mid \phi_{i}\bigr)}
     {P\bigl(x,\,v,\,Z^{(t)} \mid \phi_{i}\bigr)}.
\end{equation}
Since the denominator $P(x,\,v,\,Z^{(t)} \mid \phi_{i})$ does not depend on $\theta$, we consider the proportional relationship instead.

We now assume conditional independence of the text $x$, numerical data $v$, and previous responses $Z^{(t)}$ given the stance $\theta$:
\begin{equation}
\label{eq:proof_extended_independence}
P\bigl(x,\,v,\,Z^{(t)} \mid \theta,\,\phi_{i}\bigr)
=
P\bigl(x,\,v \mid \theta,\,\phi_{i}\bigr)\,
P\bigl(Z^{(t)} \mid \theta,\,\phi_{i}\bigr).
\end{equation}
Substituting this into the numerator of Equation~\eqref{eq:proof_extended_posterior_start}, we get:
\begin{align}
    P\bigl(x,\,v \mid \theta,\,\phi_{i}\bigr)\,P\bigl(Z^{(t)} \mid \theta,\,\phi_{i}\bigr)\,P\bigl(\theta \mid \phi_{i}\bigr).
\end{align}

Furthermore, we assume that each element $z_{j}^{(t)}$ in $Z^{(t)} = \{z_{j}^{(t)}\}_{j=1}^{n}$ is conditionally independent given the same belief $\theta$:
\begin{equation}
\label{eq:proof_extended_product}
P\bigl(Z^{(t)} \mid \theta,\,\phi_{i}\bigr)=\prod_{j=1}^{n}
P\bigl(z_{j}^{(t)} \mid \theta,\,\phi_{i}\bigr).
\end{equation}
Substituting into Equation~\eqref{eq:proof_extended_posterior_start}, we obtain:
\begin{align*}
P\bigl(\theta \mid x,\,v,\,Z^{(t)},\,\phi_{i}\bigr)
&\propto
P\bigl(x,\,v,\,Z^{(t)} \mid \theta,\,\phi_{i}\bigr)\,
P\bigl(\theta \mid \phi_{i}\bigr) \\
&=
\left[
  P\bigl(x,\,v \mid \theta,\,\phi_{i}\bigr)\,
  P\bigl(Z^{(t)} \mid \theta,\,\phi_{i}\bigr)
\right]\,
P\bigl(\theta \mid \phi_{i}\bigr) \\
&=
P\bigl(x,\,v \mid \theta,\,\phi_{i}\bigr)\,
P\bigl(\theta \mid \phi_{i}\bigr)\,
\prod_{j=1}^{n}
P\bigl(z_{j}^{(t)} \mid \theta,\,\phi_{i}\bigr).
\end{align*}
This corresponds to the posterior decomposition in Equation~\eqref{eq:lemma42_posterior_extended} of Lemma~\ref{lemma2_extended}.

Therefore, under Assumption~\ref{eq:assumption_extended}, both Lemma~\ref{lemma1_extended} and Lemma~\ref{lemma2_extended} hold.
\end{proof}

\end{document}